# Interpretation of Mott-Schottky Plots of Photoanodes for Water Splitting

Sandheep Ravishankar[1*], Juan Bisquert[2] and Thomas Kirchartz[1,3]

[1]IEK-5 Photovoltaik, Forschungszentrum Jülich, 52425 Jülich, Germany
[2]Institute of Advanced Materials, Universitat Jaume I, Castellón de la Plana 12071, Spain
[3]Faculty of Engineering and CENIDE, University of Duisburg-Essen, Carl-Benz-Str. 199, 47057 Duisburg, Germany

*author for correspondence, email: s.ravi.shankar@fz-juelich.de

**Abstract**
A large body of literature reports that both bismuth vanadate and haematite photoanodes are semiconductors with an extremely high doping density between $10^{18} - 10^{21}$ cm$^{-3}$. Such values are obtained from Mott-Schottky plots by assuming that the measured capacitance is dominated by the capacitance of the depletion layer formed by the doping density within the photoanode. In this work, we show that such an assumption is erroneous in many cases because the injection of electrons from the collecting contact creates a ubiquitous capacitance step that is very difficult to distinguish from that of the depletion layer. Based on this reasoning, we derive an analytical resolution limit that is independent of the assumed active area and surface roughness of the photoanode, below which doping densities cannot be measured in a capacitance measurement. We find that the reported doping densities in literature lie very close to this value and therefore conclude that there is no credible evidence from capacitance measurements that confirms that bismuth vanadate and haematite photoanodes contain high doping densities.

**Introduction**
Bismuth vanadate (BiVO$_4$) and haematite (α-Fe$_2$O$_3$) semiconductor films have been intensely studied over the last two decades as viable photoanodes for efficient solar water oxidation. Such investigations have mainly involved design and material modifications, such as nanostructuring,[1-3] addition of charge selective or passivating layers,[4-7] doping,[8, 9] thermal treatments that alter their bulk and surface properties,[10, 11] and the use of catalysts[12-14] that maximise charge transfer from the photoanode to the electrolyte. However, while the photoconversion efficiencies of these devices have improved significantly, a deeper understanding of their physics of operation is still lacking. This is because of the limitations of the interpretative framework used to analyse the results of commonly-used characterization techniques such as impedance spectroscopy (IS), intensity-modulated photocurrent spectroscopy (IMPS) and Mott-Schottky analysis ($C^{-2}$ vs $V$, where $C$ is the capacitance and $V$ is the applied voltage). The interpretation of IS spectra is difficult because it requires assuming an equivalent circuit that can sometimes be quite complex and yields resistances and capacitances whose experimental evolution deviates significantly from theoretical predictions,[7, 15-17] while IMPS spectra are interpreted using the evolution of time constants[18, 19] based on a kinetic model whose validity has been questioned.[20] In the case of Mott-Schottky analysis, a large body of literature reports very high doping densities in these photoanodes between $10^{18} - 10^{21}$ cm$^{-3}$ both before and after material modification, which suggests that the BiVO$_4$ and Fe$_2$O$_3$ used in these devices are degenerate semiconductors.[21-24]

The uncertainty with regard to the Mott-Schottky analysis is particularly troubling because it implies a certain physics of operation. Since the photoanode is operated at large anodic (reverse) bias, the electric field is an important factor controlling its performance, driving electrons to the collecting contact and holes to the photoanode/electrolyte interface for subsequent transfer to the electrolyte. Such large doping densities suggest that the depletion region is substantially thinner than the photoanode at the operating voltage and the holes generated closer to the collecting contact are required to diffuse over a distance before being collected by the electric field. In such cases, bulk transport limitations can also limit the performance in addition to the more well-known slow kinetics/recombination at the photoanode/electrolyte interface.[18, 22, 25-28] Several publications have indeed questioned the interpretation of these Mott-Schottky plots and discussed the different situations under which the Mott-Schottky equation and its corresponding assumptions cannot be directly applied. Hankin et al.[29] have discussed the importance of accounting for the distribution of the applied external potential across both the depletion layer and the Helmholtz layer, especially in situations of large doping densities. Peter et al.[30] showed that the planar capacitance formula used to derive the Mott-Schottky equation is invalid for nanostructured photoanodes with spherical and cylindrical geometries especially at deep reverse (anodic) bias, where the depletion layer width deviates strongly from the $\sqrt{V_{bi} - V}$ proportionality. The maximum potential drop through the individual nanostructures is also limited by their geometry, as shown by Bisquert.[31] For nanoporous films, an additional consideration is that the electrolyte can penetrate through the thickness of the photoanode, allowing the ions in the electrolyte to shield the electric field[31] and the electrostatic potential drop to occur mainly between the substrate and the electrolyte, rather than the substrate and the nanoporous film. In fact, Fabregat-Santiago et al. showed that Mott-Schottky measurements of fluorine-doped tin oxide (FTO) substrates covered by nanoporous titanium-dioxide ($TiO_2$) films are dominated by the capacitance of the FTO substrate rather than the $TiO_2$ film.[32]

Whilst all these factors are important, the most fundamental requirement for detecting a doping or trap density in a capacitance measurement is that it needs to sufficiently affect the electrostatic potential to create a space-charge region shorter than the thickness of the semiconductor. If this does not happen, the electric field in the device remains unaffected and the charge density cannot be detected. In this work, we use this argument to determine resolution limits for the detection of charge densities of bismuth vanadate and haematite photoanodes in a capacitance measurement, previously derived in ref.[33]. This resolution limit serves as an upper limit of doping/trap densities in the material and is a function of the permittivity and thickness of the material, while being independent of the assumed active area. By analysing reported Mott-Schottky plots of haematite and bismuth vanadate photoanodes, we identify that all the reported doping/trap densities are very close to the resolution limit, indicating that they cannot credibly be assigned as actual doping densities. For any Mott-Schottky measurements in the future, we suggest that the calculated doping density be compared to the resolution limit and only if it is significantly higher than the limit, be considered an actual doping density. If not, then the value serves only as an upper limit to the actual doping density in the photoanode. Based on these results, we propose alternative band diagrams of the photoanode in different bias situations (from drift-diffusion simulations using SCAPS – a Solar Cell Capacitance Simulator)[34] with a fairly constant electric field throughout the whole photoanode, as opposed to the widely-accepted picture of a sharp potential drop at the photoanode/electrolyte interface and a field-free bulk.

**Results and Discussion**

Before analysing reported data, we first provide a brief overview of the underlying physics that allows the determination of dopant or trap densities in a Mott-Schottky plot. Figure 1(a) shows the simulated band diagram of a photoanode of thickness $d$ at equilibrium (simulation details and parameters shown in table S1 in the supporting information (SI)). An electrostatic built-in potential difference $V_{bi}$ is formed at the photoanode/electrolyte interface. The associated depletion region extends into the photoanode over a distance $w$. As shown in figure 1(b), this region is assumed to be depleted of majority carriers (hence called depletion region), with the net electrostatic potential being determined only by the density of charged dopant/trap species. In the region beyond the depletion region ($x < w$), the charge of the dopant/traps is counterbalanced by an equal density of majority carriers (in this case, electrons), leading to zero net charge and zero electrostatic potential drop, hence termed the neutral region. The depletion region can be considered a parallel-plate capacitor with capacitance

$$C = \frac{\varepsilon_r \varepsilon_0}{w}. \tag{1}$$

The solution of the Poisson equation yields the width of the depletion region as[35]

$$w = \sqrt{\frac{2\varepsilon_r \varepsilon_0}{qN_d}\left(V_{bi} + \frac{k_B T}{q} - V\right)}. \tag{2}$$

The application of a cathodic (forward) voltage reduces the electrostatic potential drop (i.e. $V_{bi} - V$ becomes smaller) in the space-charge region, while an anodic (reverse) voltage increases it. Upon the application of a small perturbation of voltage, the width of the depletion region is altered by a factor $dw$ from its steady-state value $w$ and a charge density $N_d(w) \times dw$ is swept out from the edge of the depletion region (figure 1 shows a constant doping density for simplicity but in general, the doping density is variable along the thickness of the photoanode), which manifests itself as the measured current.[36] Equation 2 implies that at deep anodic (reverse) bias, the entire thickness of the semiconductor is depleted ($w = d$) and the capacitance saturates at the geometric capacitance, given by

$$C(V \to \text{deep reverse bias}) = C_g = \frac{\varepsilon_r \varepsilon_0}{d}. \tag{3}$$

From equations 1 and 2, we obtain[36]

$$N_d(w) = -\frac{2(dC^{-2}/dV)^{-1}}{q\varepsilon_r\varepsilon_0}. \tag{4}$$

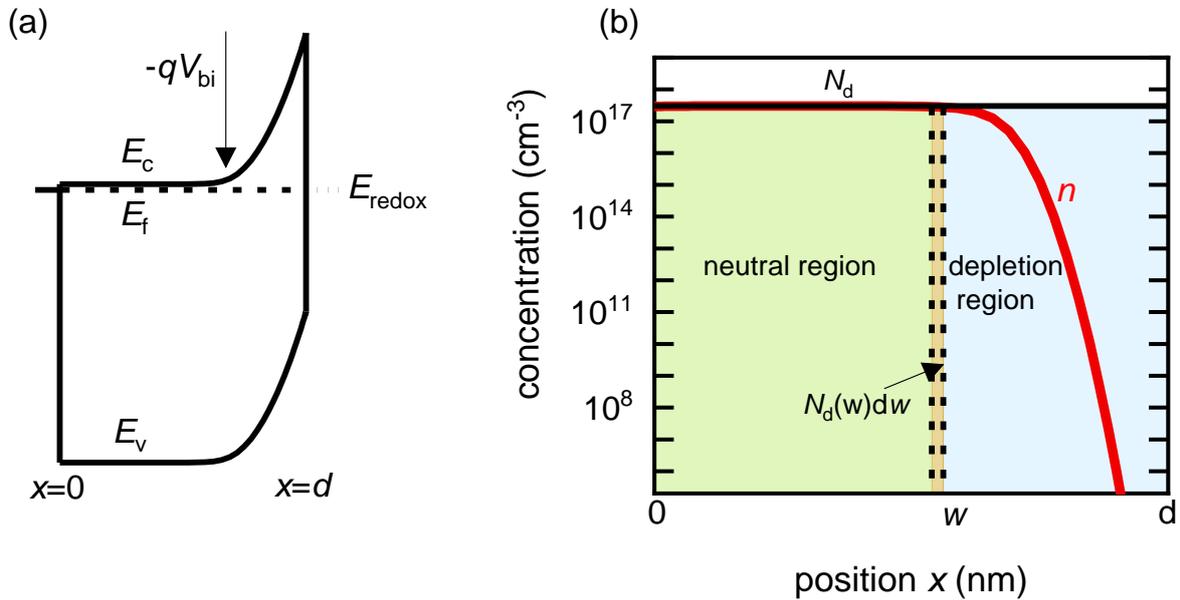

**Figure 1** (a) Simulated band diagram of a photoanode at equilibrium. The collecting contact is at $x = 0$ and the photoanode/electrolyte interface is at $x = d$. $V_{bi}$ is the built-in potential difference. (b) Carrier concentration as a function of position in a photoanode with constant doping density $N_d = 3 \times 10^{17}$ cm$^{-3}$. When a small voltage perturbation is applied, a charge density $N_d(w) \times dw$ is swept out from the edge of the depletion region of width $w$ (yellow region). Simulation parameters are provided in table S1 in the SI.

Equation 4 implies that the inverse slope of the Mott-Schottky plot at a given voltage is proportional to the doping density at the edge of the depletion region width $w$ at the same voltage. Therefore, equation 4 can be plotted against $w = \varepsilon_r \varepsilon_0 / C(V)$ (termed the 'profiling distance') (equation 1) to determine the spatial distribution of dopant/trap densities in the photoanode (termed a 'doping profile' plot. Examples of doping profiles are provided in ref.[33]). An alternate practice commonly used in the solar fuel community is to fit a straight line through the $C^{-2}$ vs $V$ data to obtain an average dopant/trap density. The intercept of this line on the $V$ axis also allows determining the flatband potential $V_{fb}$, corresponding to the region of capacitance saturation at large cathodic bias.

The correct application and analysis of the Mott-Schottky method requires acknowledging some fundamental resolution limits inherent to it, based on simple electrostatics. For a density of charge to be detected in a capacitance measurement, it must affect the electrostatic potential sufficiently. In the case of a non-porous semiconductor layer of thickness $d$ sandwiched between two electrodes, this charge density must then be significantly larger than the surface charge density $\sigma$ on the electrodes. This leads to the condition[37]

$$qN_d d > \sigma, \qquad (5)$$

where $N_d$ is a charge density per unit volume inside the semiconductor layer. A photoanode can be approximated as such a device because it employs a highly-doped semiconductor (such as FTO) to collect the electrons on the substrate side, while at the opposite side, the electrolyte can be considered a reservoir of charge with a large capacitance, akin to a metal (this reasoning was used in the simulations to model the photoanode, see section A1 in the SI). Equation 5 can also be restated as – the depletion layer width must be shorter than the thickness of the semiconductor film i.e. $d > w$. An alternate way of representing this argument is that the Debye length $L_D$, which is the distance over which the electrostatic potential drops by $k_B T/q$, is much smaller than the thickness $d$ of the semiconductor.[37] Therefore, if the charge density $N_d$ is not high enough to modify the electrostatic potential, a constant electric field is maintained through the absorber layer and only the charge density on the electrodes is measured, based on the theoretical development in equations 1-4.

The subsequent question that arises based on the discussion so far is: in situations where the doping/trap density does not affect the capacitance, what generates the capacitance step in a Mott-Schottky plot that is ubiquitously observed? In order to answer this question, it is important to first establish what a forward and reverse bias is for the photoanode. The definition of a forward bias is unclear since the applied voltage at the collecting contact is referenced to an external redox potential (three-electrode configuration), usually a Ag/AgCl electrode. This means that the voltage applied to the absorber layer is unknown since we do not have access to the hole Fermi level from the electrolyte side via a metal contact (see figure 1(a)). However, we can define a forward bias using the relative change in the electron Fermi level position (that we have access to from the collecting contact side) from an equilibrium to a non-equilibrium steady-state situation by assuming that the hole Fermi level at the absorber/electrolyte interface (i.e. at $x = d$ in figure 1(a)) is pinned to the redox level of the electrolyte. Thus, for a Mott-Schottky measurement, we have the total forward bias as

$$V_{forward} = V_{OCP,dark} - V, \qquad (6)$$

where $V_{OCP,dark}$ is the dark open-circuit potential and $V$ is the applied potential. Note that both $V_{ocp,dark}$ and $V$ are measured with respect to the same reference electrode potential $V_{ref}$. A positive forward bias implies the application of a cathodic potential that leads to injection of electrons into the absorber layer from the metal contact, while a reverse bias is a negative value of $V_{forward}$ that corresponds to large anodic potentials. The maximum forward bias $V_{max,MS}$ in a Mott-Schottky plot is then given by

$$V_{\text{max,MS}} = V_{\text{OCP,dark}} - V_{\text{FB,app}}, \qquad (7)$$

where $V_{\text{FB,app}}$ is the apparent flatband potential, obtained from the intercept of the linear Mott-Schottky region on the voltage axis. Figure 2(a) shows cyclic voltammetry curves of a bismuth vanadate photoanode, both in the dark and under illumination. Based on equation 6 and knowledge of the dark open-circuit potential, we can then define the regions in the cyclic voltammetry curve that correspond to forward and reverse bias, shown using dashed lines in figure 2(a).

We now proceed to discuss the origin of the capacitance step observed in a Mott-Schottky measurement. Figures 2(b) to 2(e) show simulated band diagrams (by numerically solving the drift-diffusion equations, see section A1 for description of simulations and section A2 for discussion of parameters in the SI) of a doped photoanode in the dark under different bias conditions, from reverse bias to equilibrium, then forward bias and ultimately flatband. We note that the simulations are shown using the same potential scale defined in equation 6, where the dark open-circuit potential is zero volts since no reference electrode is considered for the simulations. Due to the large doping density in the photoanode, there is a sharp potential drop at the photoanode/electrolyte interface, creating a small depletion layer and a large neutral region. This depletion region is eventually reduced in width and ultimately removed upon application of a forward bias. The corresponding change in net charge density (in this case, electron density, since the hole density is several orders of magnitude lower) is shown in figure 2(f). The simulated capacitance-voltage behaviour and the corresponding Mott-Schottky plot are shown in figure 2(g). The upward step in capacitance moving from anodic to cathodic voltage creates a linear Mott-Schottky region (dashed lines in figure 2(g)), as expected for a depletion capacitance. While the capacitance behaviour is obtained from a simulated small-perturbation capacitance-voltage measurement, we find that the ratio of the differential steady-state net charge density and differential voltage (i.e. $C = \mathrm{d}Q/\mathrm{d}V$) provides an accurate estimation of the capacitance obtained from the small-perturbation measurement (see figure S1(a) in the SI).

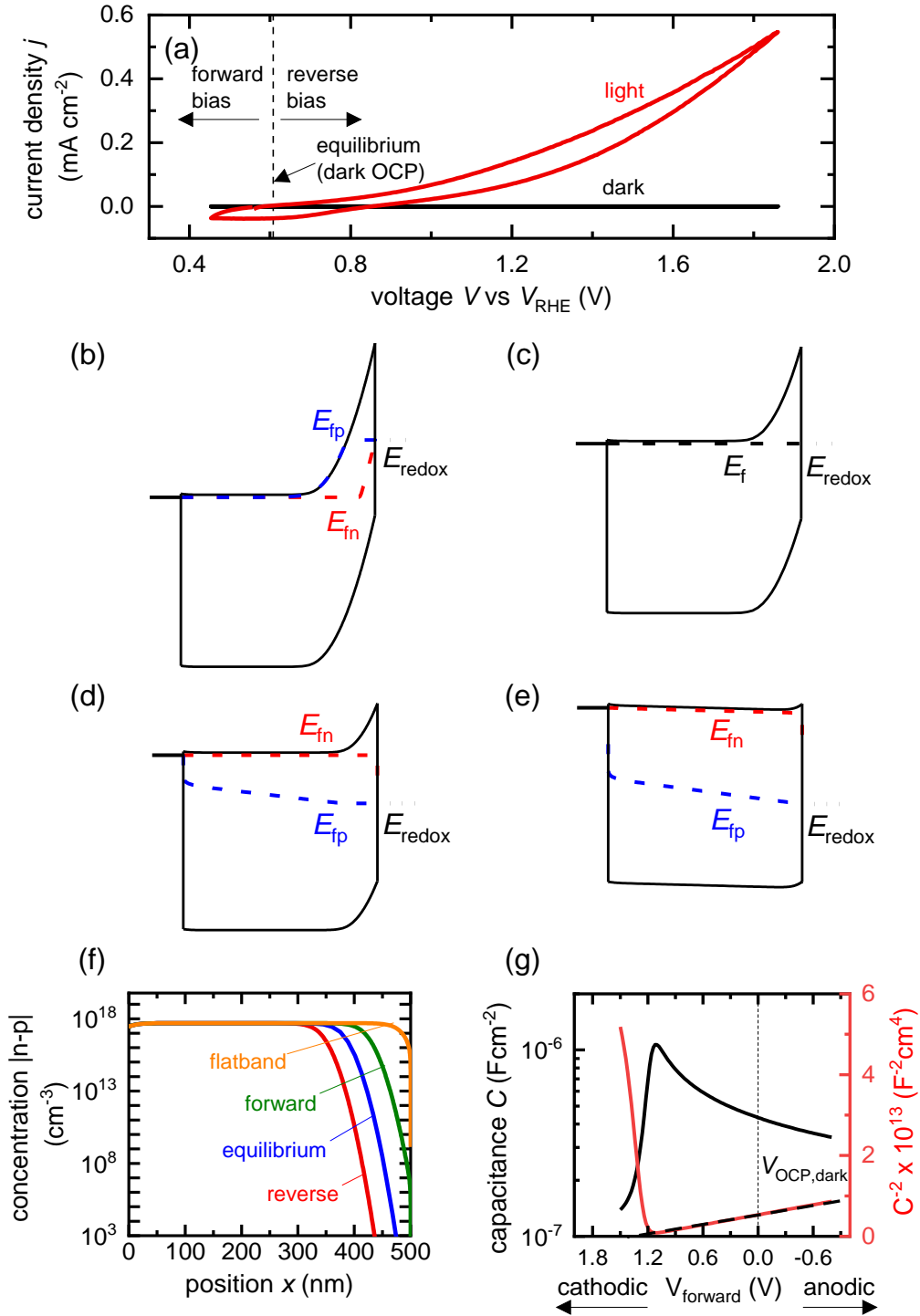

**Figure 2** (a) Cyclic voltammetry curves of a $BiVO_4$ photoanode in the dark and under one sun illumination. Also shown are the voltages corresponding to large forward bias, equilibrium (dark open-circuit potential) and deep reverse bias (anodic voltages). Simulated band diagrams of a doped $BiVO_4$ photoanode ($N_d = 5 \times 10^{17}$ cm$^{-3}$) in the dark at (b) reverse bias, (c) equilibrium, (d) forward bias and (e) flatband conditions. (f) shows the net concentration $|n-p|$ in the photoanode as a function of position for the different bias situations in (b)-(e). The net concentration is dominated by the electron density, which is several orders of magnitude larger than the hole density in all cases. (g) Capacitance step and corresponding Mott-Schottky plot (dashed line shows linear Mott-Schottky region) from a simulated small-perturbation capacitance-voltage measurement ($10^3$ Hz). This capacitance corresponds to the depletion capacitance. Simulation parameters are shown in Table S2 in the SI.

We now consider an undoped (intrinsic) photoanode, whose band diagrams corresponding to different bias situations in the dark are shown in figures 3(a) to 3(d). These situations are identical to that in figures 2(b) to 2(e) except for the fact that there is no depletion layer in the photoanode. Instead, there is a constant electric field through the thickness of the semiconductor at equilibrium, determined by the difference between the electron contact workfunction and the energetic distance of the redox level from the surface vacuum level. Upon applying a forward bias, the electric field and the tilting of the bands are reduced. This is the situation shown in figure 3(c) where the electric field is smaller compared to figure 3(b). Upon applying a larger forward bias, the electric field nearly vanishes and the flatband condition shown in figure 3(d) is reached. An undoped photoanode also shows a large increase in net charge density through its thickness upon the application of a forward bias, shown in figure 3(e), resulting in the capacitance step in figure 3(f) (the differential capacitance calculated from the steady-state net charge density in figure 3(e) provides a decent approximation of the capacitance, see figure S1(b) in the SI). We note that in the case of the undoped photoanode, the injected charge that causes the capacitance step is that of the majority carriers (electrons) and not the minority carriers (holes). This occurs due to the difference in injection barriers for electrons and holes at the electron-contact/photoanode interface and photoanode/electrolyte interface, 0.05 eV and 1.05 eV respectively (see table S3 in the SI). Therefore, the equilibrium concentrations for electrons and holes at the electron-contact/photoanode and photoanode/electrolyte interfaces respectively are $n_0 = 2.9 \times 10^{17}$ cm$^{-3}$ and $p_0 = 5.78$ cm$^{-3}$. The electron and hole concentrations along the length of the photoanode at equilibrium is then given by

$$n(x) = n_0 \exp\left(\frac{-qFx}{k_BT}\right), \tag{8}$$

$$p(x) = p_0 \exp\left(\frac{qFx}{k_BT}\right). \tag{9}$$

The electric field $F$ is given by

$$F = -\frac{(V_{bi}-V)}{d}, \tag{10}$$

where $V_{bi}$ is the built-in voltage through the thickness of the intrinsic photoanode and $V$ is the applied voltage. Thus, the electron and hole concentrations increase exponentially with applied forward voltage from their equilibrium concentrations. Since $n_0 \gg p_0$, the photoanode contains an excess of electrons (pseudo n-type, as seen from figures 3(c), (d) and (e)) even though it is an intrinsic semiconductor. This capacitance can be labelled as an electrochemical capacitance, since it involves a change in both the electrostatic potential and electron Fermi level to generate the exponential increase in charge density. This capacitance step makes an apparent linear region (dashed lines in figure 3(f)) in a Mott-Schottky representation, which can easily be mistaken as a signature of a real doping density. We note that in cases where the depletion capacitance is overshadowed by the electrochemical capacitance at large forward bias, the Mott-Schottky region can in theory be observed and also fitted at large anodic (deep reverse) biases, as shown in figure S2 in the supporting information. Additionally, the apparent built-in voltage obtained from this plot is very similar to that generated by the depletion capacitance (figure 2(g)), while both cases possess the same net built-in electrostatic potential drop at equilibrium that corresponds to the difference between the electron contact workfunction and the energetic distance of the water oxidation redox potential to the surface vacuum level. This effect has been discussed in literature[38] and is related to the fact that at potentials close to flatband, the electrochemical capacitance saturates at a maximum value because the net change in electron density is cancelled out by an equivalent change in hole density at each spatial location.[39]

We have so far considered two extreme cases, the first one being a highly-doped semiconductor where the capacitance response is dominated by the depletion capacitance (figure 2) and the second one being an undoped, intrinsic photoanode where the capacitance response is dominated by the electrochemical capacitance (figure 3). However, there are several reports in literature suggesting the presence of a background doping density in these photoanodes, arising from oxygen vacancies and hydrogen donors in the case of bismuth vanadate[40-42] and oxygen vacancies in the case of haematite.[43-45] We therefore simulate an intermediate case where a doping density exists in the photoanode but the capacitance step is still dominated by the electrochemical capacitance (see figure S3). In this scenario, the doping density is not high enough to modify the electrostatic potential and create a depletion region, which leads to a similar situation as the undoped case in figure 3. Thus, the electrochemical capacitance can dominate the total capacitance response even in situations where a doping density is present in the photoanode. Additionally, since capacitance measurements rely on the amount of charge placed on the dopants (that create the depletion region and the subsequent capacitance response), positively-charged and negatively-charged dopants can cancel each other out and remain invisible in these methods. To provide an intuitive understanding of the capacitance step generated by the injection of electrons from the electron contact, we show a schematic of the process in figure 4.

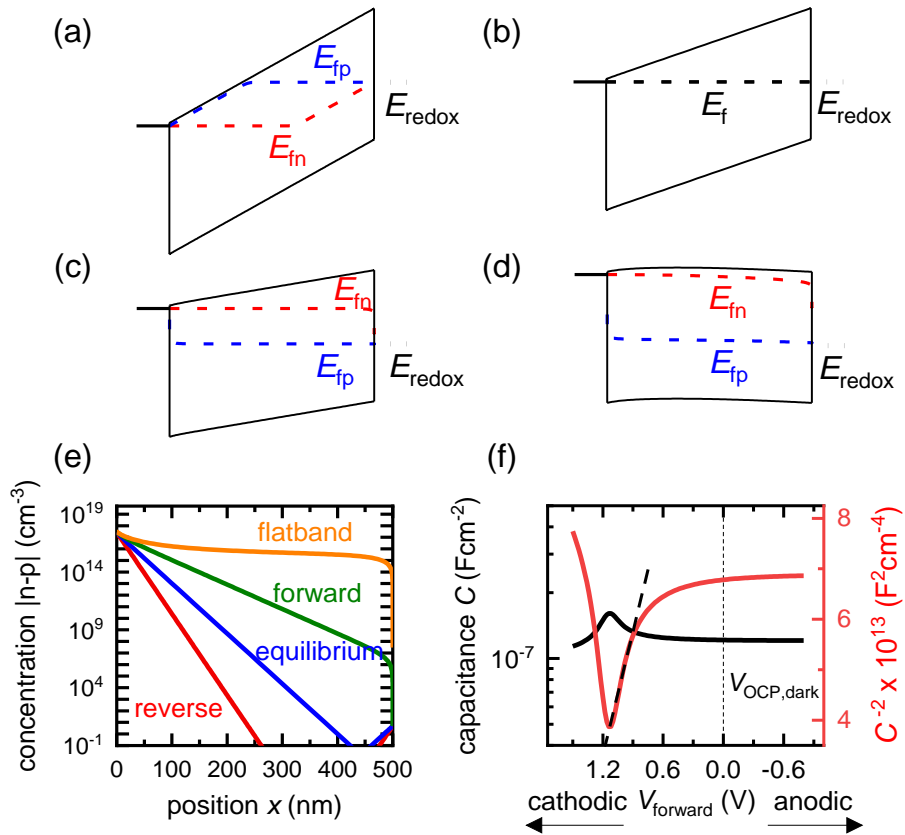

**Figure 3** Simulated band diagrams of an undoped (intrinsic) photoanode in the dark at (a) deep reverse bias, (b) equilibrium, (c) forward bias and (d) flatband conditions. (e) shows the corresponding total concentration $|n-p|$ as a function of position in the photoanode for the different situations in (a)-(d). The hole density is several orders lower in magnitude than that of the electrons in all cases, which means the total charge density in the device is dominated by the electron density (majority carriers). (f) shows the capacitance step and corresponding Mott-Schottky plot and apparent flatband potential from a simulated small-perturbation capacitance-voltage measurement ($10^3$ Hz). Simulation parameters are shown in Table S3 in the SI.

To identify if the capacitance associated to injection of electrons from the metal contact is indeed a significant factor affecting the Mott-Schottky measurements, we calculate the magnitude of forward bias used in typical Mott-Schottky measurements of bismuth vanadate and haematite photoanodes reported in literature, shown in figure 5(a). This figure shows that most of the Mott-Schottky plots are measured under at least a few hundred mV of forward bias, indicating that charge injection is a process that can significantly affect the capacitance response in a Mott-Schottky measurement.

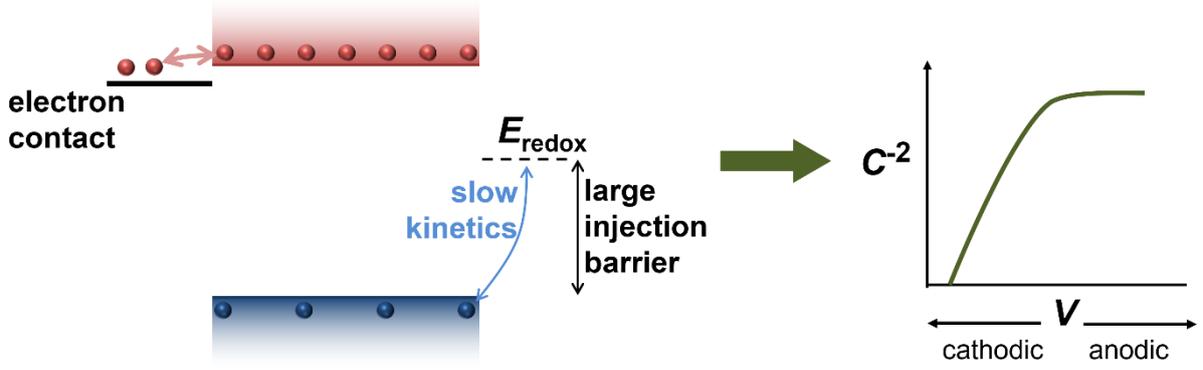

**Figure 4** Schematic of the process of injection of electrons (red spheres) from the electron contact into the photoanode that generates the capacitance step in the dark and consequent Mott-Schottky plot observed experimentally. The hole density (blue spheres) in the photoanode is very low due to the large injection barrier at the photoanode/electrolyte interface, causing the electrons to dominate the capacitance step.

To determine the form of this capacitance, we consider that in case of a field-free, intrinsic photoanode with symmetric injection barriers, we have $n = p$ which gives $n, p \propto \exp(qV/2k_BT)$ while in the case of a doped layer, we have the minority carriers taking the entire voltage, yielding $p \propto \exp(qV/k_BT)$. Since the electrochemical capacitance is proportional to the injected charge density, we can describe the capacitance evolution versus voltage as

$$C(V) = C_g + C_0 \exp\left(\frac{qV}{m_{CV} k_B T}\right), \quad (8)$$

where $C_g$ is the geometric capacitance and $m_{CV}$ is a factor that controls the slope of the exponential capacitance transition. Carrying out a Mott-Schottky analysis of equation 8, we obtain the doping density as[33]

$$N_{d,min} = \frac{27 m_{CV} k_B T \varepsilon_r \varepsilon_0}{4 q^2 d^2}, \quad (9)$$

where $\varepsilon_r$ and $\varepsilon_0$ are the relative permittivity of the photoanode and vacuum permittivity respectively and $d$ is the thickness of the photoanode.

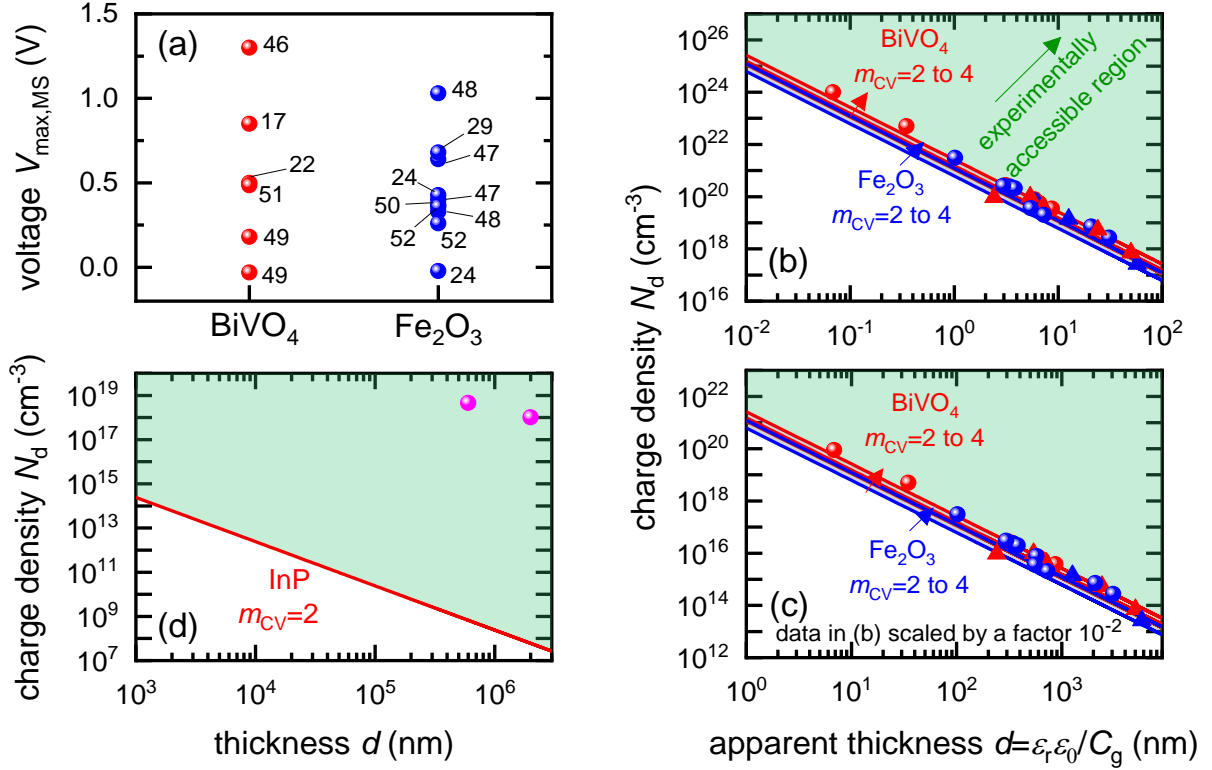

**Figure 5** (a) Literature data points (corresponding to reference number as shown) of the maximum forward voltage applied in a Mott-Schottky measurement calculated from the difference between the dark open-circuit voltage and the flatband potential, see equation 7. Cathodic (negative) voltages imply the exponential injection of electrons into the photoanode from the collecting contact during a Mott-Schottky measurement. (b) Doping densities of $BiVO_4$ (red) and $Fe_2O_3$ (blue) photoanodes as a function of apparent device thickness, calculated from Mott-Schottky plots reported in literature. Triangles indicate data points where the capacitance saturates at deep reverse (anodic) bias. The minimum doping density resolvable from capacitance measurements for different $m_{CV}$ values (see equation 9) is also plotted. All the data points lie very close to the resolution limits and hence cannot credibly be ascribed to actual dopant densities in the device. (c) Data points in (b) plotted with the capacitance multiplied by an arbitrary factor of $10^{-2}$. The data points still remain very close to the resolution limits, indicating that the assumed active area and surface roughness factors of the device has no effect on the resolving power of the capacitance measurement. The apparent thickness was calculated from the smallest capacitance value at deep reverse (anodic) bias, see equation 3. The data points were obtained for (a) from references [17, 22, 24, 29, 46-52] and for (b) from references [1, 2, 7, 8, 17, 21-24, 44, 47, 48, 53-55]. (d) Doping densities of InP photoelectrodes as a function of reported device thickness. The doping densities were calculated from the plateau region of the doping profile, assuming a relative permittivity of $\varepsilon_r = 12.5$. The data points correspond to references [56] and [57], both of which are above the resolution limit corresponding to a valid doping density. The doping density values as obtained from the Mott-Schottky data are similar to the values provided by the suppliers of the InP single crystals (1-4 × $10^{18}$ cm$^{-3}$ and 6.7 × $10^{17}$ cm$^{-3}$ respectively). $\varepsilon_r = 68$, $\varepsilon_r = 32$ and $\varepsilon_r = 12.5$ was assumed for bismuth vanadate, haematite and indium phosphide respectively, to calculate the doping density from the plateau region of the doping profile (see figure S5 in the SI).

Equation 9 provides the apparent doping density observed when the capacitance step versus voltage occurs solely due to the injection of electrons into the photoanode and the charges on the electrode. Therefore, $N_{\text{d,min}}$ is the minimum apparent doping density that will be observed in a Mott-Schottky plot and only real doping/trap densities larger than $N_{\text{d,min}}$ can be detected in a Mott-Schottky measurement. Figure 5(b) shows reported doping/trap densities of bismuth vanadate and haematite photoanodes in literature for different thicknesses of the photoanode layer. Note that the thickness of the film is often not provided nor is it the relevant quantity in this context. The relevant effective thickness in this context is indeed the thickness derived from the capacitance plateau at large anodic (deep reverse bias) voltages using equation 3, whose validity is assumed in a Mott-Schottky measurement. The doping/trap densities are very close to the resolution limits calculated for a capacitance step with $m_{\text{CV}} = 2 - 4$ and follow the $N_{\text{d,min}} \propto d^{-2}$ proportionality in equation 9. We note that we use an $m_{\text{CV}}$ factor between 2 and 4 because figure 3 is a case of an undoped, intrinsic photoanode with an electric field, different from both the undoped, field-free intrinsic photoanode ($m_{\text{CV}} = 2$) and the doped photoanode ($m_{\text{CV}} = 1$). Thus, a simple analytical form of the capacitance cannot be achieved and the $m_{\text{CV}}$ factor can only be obtained through fitting of the rising capacitance step generated from the simulations to equation 8. This yields large $m_{\text{CV}}$ values of ~7.2 (see figure S4 in the SI), and hence we use $m_{\text{CV}}$ values between 2 and 4 for the resolution limit.

Our analysis confirms that the experimental doping densities in figure 5(b) are artefacts from a general capacitance transition of the form of equation 8 and serve as upper limits to the actual doping/trap densities in the material. Additionally, the magnitudes of the experimental doping/trap densities in figure 5(b) are very high, with most data points between $10^{19} - 10^{21}$ cm$^{-3}$. Sivula[58] has recently questioned the magnitude of these doping densities, noting that they are likely in excess of the density of states in the conduction/valence bands of these materials. Such large values are in part a consequence of overestimation of the capacitance per unit area due to consideration of the planar active area, instead of the much larger effective surface area of nanostructured and porous films. The surface roughness of such nanostructured capacitors has also been shown to significantly increase the measured capacitance.[59] This is reflected in the apparent thickness of the device in the x-axis of figure 5(b), yielding film thicknesses of (in some cases) only a few nanometres or even less.

The question then arises, whether the increase of the capacitance in nanostructured materials due to high surface areas and substantial roughness will change the position of the experimental data points relative to the validity region of the method (green region in figure 5(b)). We might assume that the capacitances are higher than the capacitances expected for flat films with a given doping concentration by a voltage-independent factor that is a consequence of surface roughness. Hence, we would have to study the situation where the capacitance is scaled with a factor < 1. We therefore multiply the capacitance data of the points in figure 5(b) by an arbitrary factor of 10$^{-2}$ and consider the effect of this multiplication both on the apparent doping density and the apparent thickness of the film. The resulting graph is shown in figure 5(c), which shows that the relative position of the experimental data and the resolution limit have not changed. The only things that have changed are the absolute values of the apparent doping density and thickness. This result is a consequence of the invariance of the resolution limit with respect to any constant multiplicative factor for the capacitance. The lines in figures 5(b) and 5(c) are lines of equal $N_{\text{d,min}} \times d^2$ (see equation 9). If we rewrite equation 9 using equations 3 and 4, we obtain

$$N_{\text{d,min}} \times d^2 \times \left(\frac{q}{2\varepsilon_r \varepsilon_0}\right) = -\frac{(dC^{-2}/dV)^{-1}}{C_g^2} = \frac{27 m k_B T}{8q}. \qquad (10)$$

Both the numerator and denominator of the middle term in equation 10 contain $C^2$ terms, which means that any scaling factors arising out of poor estimation of the active area or from surface roughness cancel out to give the constant right-hand side of equation 10. Therefore, the scaled doping densities have only moved along the $N_d \propto d^2$ line to lower magnitudes and corresponding larger apparent device thicknesses. Hence, our resolution limit is also still valid even with the use of the apparent device thickness calculated from the Mott-Schottky data, however close or far it may be to the actual device thickness. To further justify the validity of our developed resolution limit, we analyse two Mott-Schottky measurements of indium phosphide (InP) single crystal photoelectrodes, shown in figure 5(d). The data points are well above the resolution limit, corresponding to a real doping density that also closely corresponds to the values provided by the suppliers of the InP single crystals. This shows that the method is in principle working but requires the doping density to be above the detection threshold indicated in Figure 5.

In summary, we conclude that there is no significant evidence to justify that bismuth vanadate and haematite photoanodes are as highly doped as many capacitance measurements suggest. This is due to the fact that the large reported doping densities for these photoanodes are very close to the resolution limits of the capacitance-voltage method. We suggest that any doping/trap densities measured from capacitance-voltage methods in the future must be significantly larger than the resolution limit to be safely labelled as an actual doping density. In case it is close to or below the resolution limit, it can only be considered as an upper limit to the actual doping density in the photoanode. The actual upper limit of the doping density in the material can be determined only upon accurate calculation of the effective active area and surface roughness of the device.

These results provide an alternate view of the electrostatics in the device. Figure 6 shows simulated band diagrams of an intrinsic photoanode at equilibrium and at large anodic and cathodic potentials under illumination. In all cases, there is a constant electric field through the entire thickness of the photoanode. This field plays an important role in maximising the photocurrent by ensuring efficient separation and transport of electrons and holes from the bulk to the collecting contact and photoanode/electrolyte interface respectively. This is in sharp contrast to the commonly-assumed picture of a small depletion layer at the photoanode/electrolyte interface and a large, neutral bulk, which requires that the bulk electrons are required to diffuse over the entire neutral region to reach the collecting contact, while the bulk holes also diffuse across the neutral region until they are swept out by the electric field and subsequently transferred to the electrolyte. We note that due to the slow kinetics of charge transfer at the photoanode/electrolyte interface, a large density of holes must exist at this interface in order to sustain the photocurrent (see equation S4 in the SI) under illumination. This makes the hole Fermi level in figures 6(b) and 6(c) appear un-pinned at the photoanode/electrolyte interface, when in fact it makes a sharp jump down to the redox level at this point.

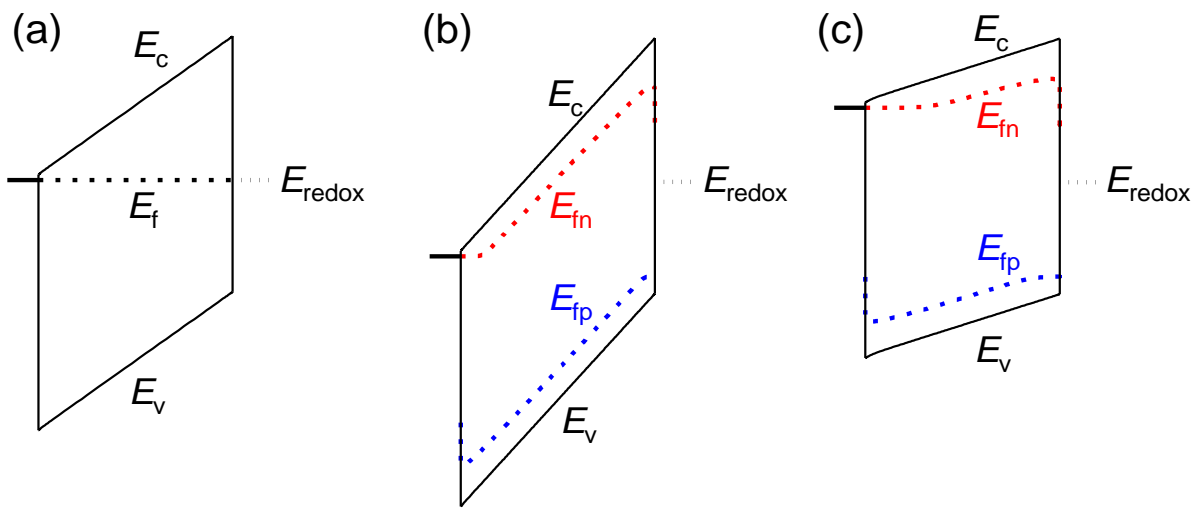

**Figure 6** Simulated band diagrams of an undoped photoanode at (a) dark equilibrium, (b) anodic potential under illumination and (c) cathodic potential under illumination. In all cases, an electric field exists throughout the thickness of the photoanode that sweeps electrons to the collecting contact and holes to the photoanode/electrolyte interface. The jump in the Fermi level of the holes in (b) and (c) at the photoanode/electrolyte interface is due to the slow kinetics of hole transfer from the valence band of the photoanode to the electrolyte. Simulation parameters are provided in table S3 in the SI.

## Conclusions

We have carried out a critical analysis of the validity of the reported large doping densities between $10^{18} - 10^{21}$ cm$^{-3}$ from Mott-Schottky measurements of haematite and bismuth vanadate photoanodes. We find that most of these Mott-Schottky measurements have been carried out at large cathodic voltages, implying significant injection of electrons from the metal contact into the photoanode. By accounting for the capacitive contribution of these injected electrons, we developed an analytical resolution limit that is independent of the assumed active area and surface roughness of the device, below which a density of dopants cannot be detected in a capacitance measurement. We identify that all of the reported doping densities in our dataset lie very close to the resolution limit, indicating that they are not actual doping densities but artefacts arising from injection of electrons into the photoanode during a Mott-Schottky measurement. These results indicate that there is no credible evidence from capacitance measurements that confirms that bismuth vanadate and haematite photoanodes contain high doping densities. We suggest that in the future, any doping/trap densities calculated from capacitance measurements should always be compared against the resolution limit. If the charge density is much larger than the limit, it can be safely labelled a doping/trap density. If it is close to or below the resolution limit, it can only be considered as an upper limit to the actual doping/trap density in the photoanode.

Based on these insights, we provide an alternate picture of the physics of operation of the photoanode, where an electric field is present throughout the thickness of the semiconductor that drives photogenerated electrons and holes from the bulk to the collecting contact and photoanode/electrolyte interface respectively, rather than a sharp potential drop at the photoanode/electrolyte interface and a large, neutral bulk region.

## Author contributions
S.R. conceptualized the project, carried out all the simulations and wrote the manuscript. J.B. and T.K. contributed to reviewing and editing the manuscript.

## Conflicts of interest
There are no conflicts of interest to declare.

## Acknowledgements
S.R. thanks Dr. Miguel García Tecedor for providing the cyclic voltammetry data in figure 2(a) and acknowledges the German Research Foundation (DFG) for the Walter-Benjamin fellowship – project number 462572437. S.R. and T.K. acknowledge funding from the Helmholtz Association via the POF-IV program. Open access publication funded by the German Research Foundation (DFG) – 491111487.

Electronic Supplementary Information

# Interpretation of Mott-Schottky Plots of Photoanodes for Water Splitting

**Sandheep Ravishankar[1*], Juan Bisquert[2] and Thomas Kirchartz[1,3]**

[1]IEK-5 Photovoltaik, Forschungszentrum Jülich, 52425 Jülich, Germany

[2]Institute of Advanced Materials, Universitat Jaume I, Castellón de la Plana 12071, Spain

[3]Faculty of Engineering and CENIDE, University of Duisburg-Essen, Carl-Benz-Str. 199, 47057 Duisburg, Germany

**A1** *Drift-diffusion Simulations*

All simulations were carried out using SCAPS – a Solar Cell Capacitance Simulator,[1] which numerically solves three coupled differential equations in one dimension. These differential equations are the Poisson equation

$$\frac{\partial^2 \varphi}{\partial x^2} = -\frac{\rho}{\varepsilon},$$ (S1)

where the space charge density is given by $\rho = q(p-n-N_A+N_D)$, where $n$ and $p$ are the densities of electrons and holes, $N_A$ is the density of ionized acceptor-like defects and $N_D$ the density of ionized donor-like defects. At steady state ($dn/dt = 0$ and $dp/dt = 0$), the continuity equations are given by

$$-\frac{1}{q}\frac{dJ_n(x)}{dx} = -D_n\frac{d^2 n(x)}{dx^2} - F\mu_n\frac{dn(x)}{dx} = G(x) - R(x,n,p)$$ (S2)

for electrons and

$$\frac{1}{q}\frac{dJ_p(x)}{dx} = -D_p\frac{d^2 p(x)}{dx^2} + F\mu_p\frac{dp(x)}{dx} = G(x) - R(x,n,p)$$ (S3)

for holes. In equations S2 and S3, $D_{n,p} = kT\mu_{n,p}/q$ are the diffusion coefficients for electrons and holes, $\mu_{n/p}$ are the mobilities of electrons and hols, $R$ is the recombination rate, $G$ is the generation rate and $F$ is the electric field. Equations S1 to S3 are numerically solved using suitable boundary conditions for each differential equation. The electrolyte was modelled as a metallic contact whose Fermi level corresponds to the water oxidation potential (+1.23 V vs RHE). The electron and hole currents at the respective interfaces with the collecting contact and electrolyte are modelled using the equation

$$j = k(n - n_0),$$ (S4)

where $k$ is a charge transfer velocity and $n$ and $n_0$ are the non-equilibrium and equilibrium electron/hole densities respectively. The slow kinetics of holes at the photoanode/electrolyte interface was thus modelled by choosing a low value of $k$ for the holes at this interface (see table S1, S2 and S3). The voltage was applied to the electron contact. Further information on drift-diffusion simulations can be found in references [2-4].

**Table S1** Parameters used for the SCAPS simulations in figure 1.

| parameter | electron contact | photoanode (Bismuth Vanadate) | electrolyte |
|---|---|---|---|
| thickness (nm) | | 500 nm | |
| relative permittivity | | 68 | |
| bandgap (eV) | | 2.4 | |
| electron affinity (eV) | | 4.15 | |
| effective DOS CB ($cm^{-3}$) | | $2 \times 10^{18}$ | |
| effective DOS VB ($cm^{-3}$) | | $2 \times 10^{18}$ | |
| radiative recombination coefficient ($cm^3/s$) | | $6 \times 10^{-11}$ | |
| injection barrier (eV) | 0.05 | | 1.05 |
| electron mobility ($cm^2/Vs$) | | 0.02 | |
| hole mobility ($cm^2/Vs$) | | 0.02 | |
| doping density ($cm^{-3}$) | | $3 \times 10^{17}$ | |
| workfunction (eV) | 4.2 | | 5.5 |
| electron charge transfer velocity (cm/s) | $10^7$ | | $10^{10}$ |
| hole charge transfer velocity (cm/s) | $10^{10}$ | | $10^1$ |

**Table S2** Parameters used for the SCAPS simulations in figure 2.

| parameter | electron contact | photoanode (Bismuth Vanadate) | electrolyte |
|---|---|---|---|
| thickness (nm) | | 500 nm | |
| relative permittivity | | 68 | |
| bandgap (eV) | | 2.4 | |
| electron affinity (eV) | | 4.15 | |
| effective DOS CB ($cm^{-3}$) | | $2 \times 10^{18}$ | |
| effective DOS VB ($cm^{-3}$) | | $2 \times 10^{18}$ | |
| radiative recombination coefficient ($cm^3/s$) | | $6 \times 10^{-11}$ | |
| injection barrier (eV) | 0.05 | | 1.05 |
| electron mobility ($cm^2/Vs$) | | 0.02 | |
| hole mobility ($cm^2/Vs$) | | 0.02 | |
| doping density ($cm^{-3}$) | | $5 \times 10^{17}$ | |
| workfunction (eV) | 4.2 | | 5.5 |
| electron charge transfer velocity (cm/s) | $10^7$ | | $10^{10}$ |
| hole charge transfer velocity (cm/s) | $10^{10}$ | | $10^1$ |

**Table S3** Parameters used for the SCAPS simulations in figure 3.

| parameter | electron contact | photoanode (Bismuth Vanadate) | electrolyte |
|---|---|---|---|
| thickness (nm) | | 500 nm | |
| relative permittivity | | 68 | |
| bandgap (eV) | | 2.4 | |
| electron affinity (eV) | | 4.15 | |
| effective DOS CB (cm$^{-3}$) | | $2 \times 10^{18}$ | |
| effective DOS VB (cm$^{-3}$) | | $2 \times 10^{18}$ | |
| radiative recombination coefficient (cm$^3$/s) | | $6 \times 10^{-11}$ | |
| injection barrier (eV) | 0.05 | | 1.05 |
| electron mobility (cm$^2$/Vs) | | 0.02 | |
| hole mobility (cm$^2$/Vs) | | 0.02 | |
| doping density (cm$^{-3}$) | | 0 | |
| workfunction (eV) | 4.2 | | 5.5 |
| electron charge transfer velocity (cm/s) | $10^7$ | | $10^{10}$ |
| hole charge transfer velocity (cm/s) | $10^{10}$ | | $10^1$ |

**A2 Discussion of the parameters**

**Bandgap:** Since a BiVO$_4$ photoanode was considered, the bandgap was chosen from ref.[5].

**Relative permittivity:** The relative permittivity of BiVO4 was chosen from ref.[6].

**Electron affinity and injection barriers:** The electron affinity was chosen based on the workfunction of the metal contact, to make a small injection barrier for the electrons at the electron contact. The hole injection barrier (distance between valence band and water oxidation redox level) was chosen from ref [7].

**Density of states:** The density of states was chosen arbitrarily.

**Radiative recombination coefficient:** The radiative recombination coefficient was chosen arbitrarily.

**Workfunction:** The electron contact was assumed to be Ag and hence, a workfunction of 4.2 eV was chosen. The energetic distance of the water oxidation redox level to surface vacuum level was chosen based on ref [7].

**Mobility:** The electron mobility was reduced by one order from that reported in ref.[8] to account for the fact that we consider an undoped BiVO$_4$ photoanode. We assumed electron and hole mobilities to be the same for simplicity. For the doped BiVO$_4$ simulations, we assume that the mobilities remain the same as that of the undoped case for simplicity.

**Charge transfer velocities:** For the electron and hole transfer rates, we assume that the interfaces are blocking for the opposite carrier. This means that all the excess holes at the electron contact/photoanode interface recombine and are lost, and all the excess electrons at the photoanode/electrolyte interface recombine and are lost. Thus, the charge transfer velocity for holes at the electron-contact/photoanode interface and for electrons at the photoanode/electrolyte interface was set very high, to a value of $10^{10}$ cm/s. We also assumed that the metal contact is a good extractor of electrons and hence set the electron charge transfer velocity at this interface to $10^7$ cm/s. For the hole transfer velocity from the photoanode to the electrolyte, we set the value very low – to $10^1$ cm/s to account for the slow kinetics of hole transfer.[9-12] While the value itself is not obtained from literature, it was chosen demonstratively to simulate the band diagram under such a condition of slow hole transfer.

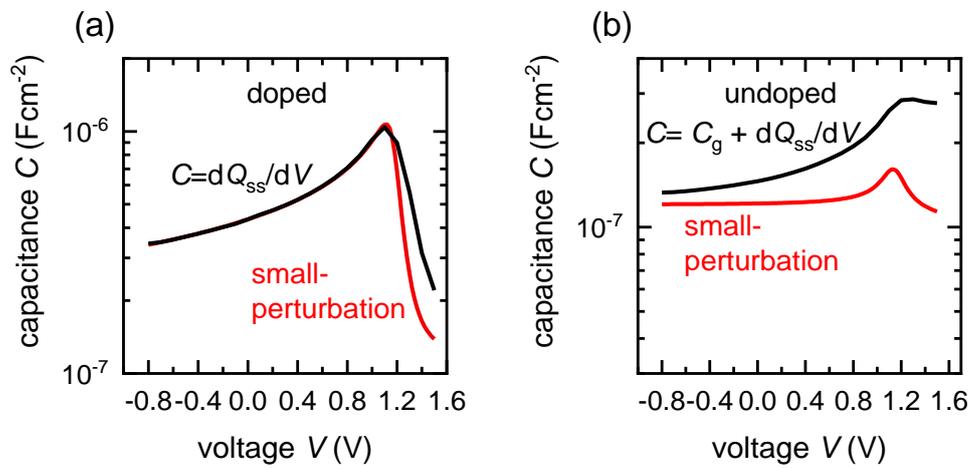

**Figure S1** Capacitance calculated from SCAPS simulations using two methods – from a small-perturbation of voltage (small-perturbation) and from the ratio of the differential of the average net charge density $Q_{ss}$ in the photoanode at steady-state and the differential voltage. The average net steady-state charge density is calculated as $Q_{ss} = qd(\frac{1}{d})\int_0^d |n - p|\mathrm{d}x$.

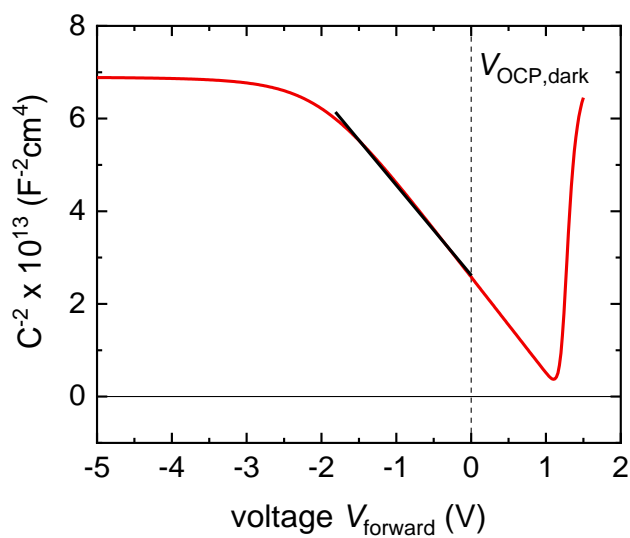

**Figure S2** Simulated Mott-Schottky behaviour of a doped photoanode ($N_d = 10^{17}$ cm$^{-3}$) in the dark, with the same parameters as table S1. In this situation, the capacitance at equilibrium and deep reverse bias is dominated by the depletion capacitance, which means the linear Mott-Schottky region should be clearly visible at (and hence can be fitted at) deep reverse or anodic biases, as shown using the black line.

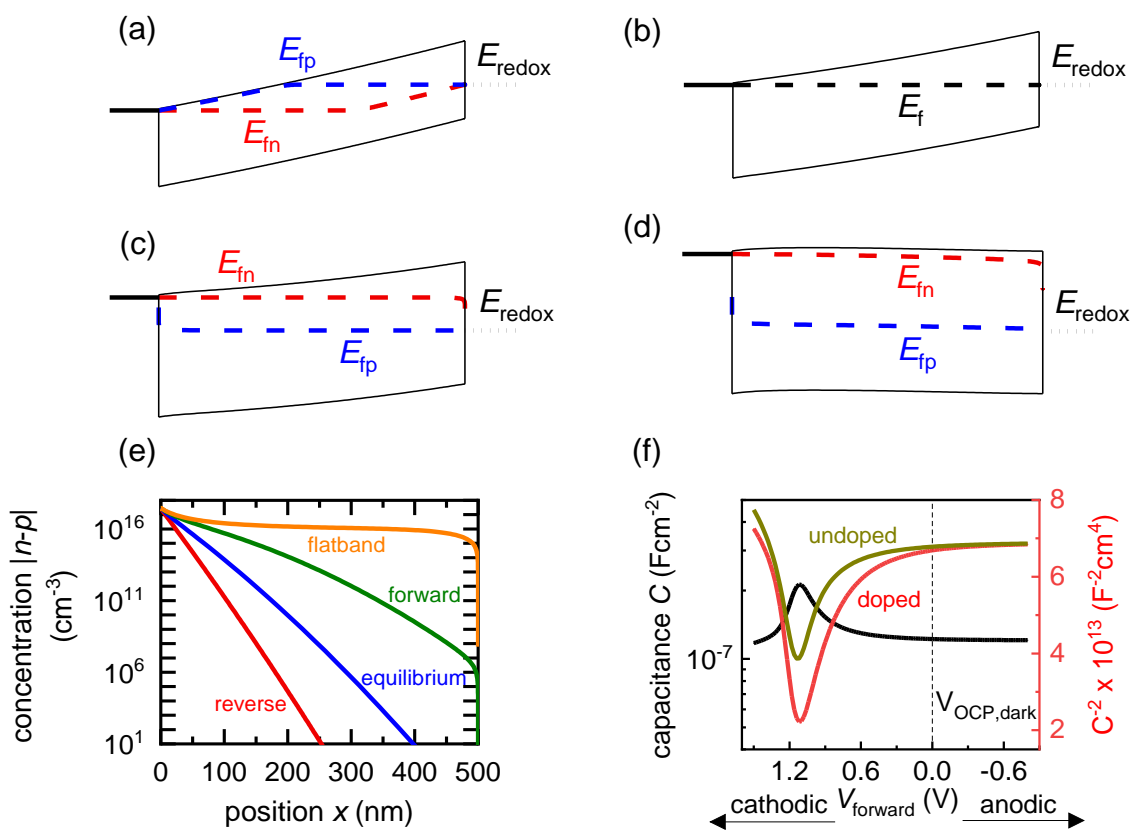

**Figure S3** Simulated band diagrams of a doped photoanode ($N_d = 10^{16}$ cm$^{-3}$) in the dark at

(a) deep reverse bias, (b) equilibrium, (c) forward bias and (d) flatband conditions. (e) shows the corresponding total concentration |n-p| as a function of position in the photoanode for the different situations in (a)-(d). (f) shows the capacitance step and corresponding Mott-Schottky plot from a simulated small-perturbation capacitance-voltage measurement ($10^3$ Hz). Also shown in (f) is the Mott-Schottky plot generated by an undoped (intrinsic) absorber layer. Simulation parameters are shown in Table S1. The Mott-Schottky plot due to the existence of a doping density is still very difficult to distinguish experimentally from that generated by an undoped absorber layer.

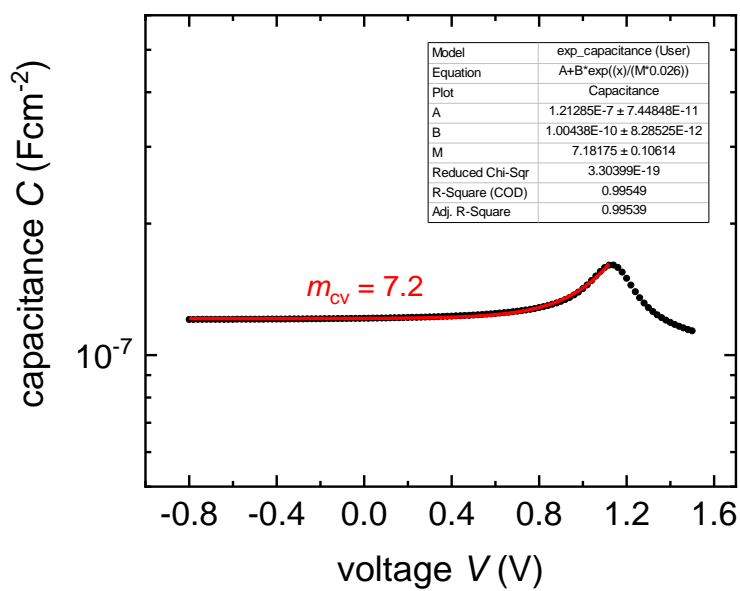

**Figure S4** Fitting (red line) of the rising capacitance step generated by an undoped, intrinsic photoanode (figure 3(f) in the main text) to equation 8 in the main text. The $m_{CV}$ factor obtained from the fit is also shown.

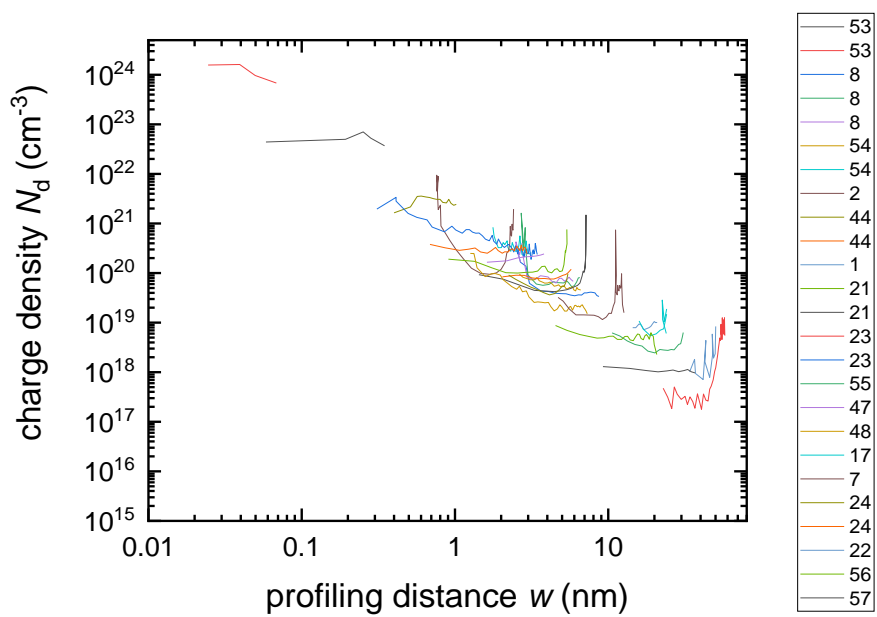

**Figure S5** Calculated doping profiles of the literature data points in figures 4(b) and 4(d). A relative permittivity value of 32 for haematite, 68 for bismuth vanadate and 12.5 for indium phosphide was assumed to calculate the doping profiles. The doping densities were obtained from the plateau regions of the doping profiles. The numbers of the references in the main paper corresponding to these doping profiles are shown in the label.